\documentclass[aps,reprint,10pt,notitlepage,pra,
               onecolumn,
               amsfonts,amssymb,amsmath,
               showpacs,
               nobibnotes,
               nofootinbib,
               ]{revtex4-1}
\usepackage{graphicx}
\usepackage{amssymb}
\usepackage{amstext}
\usepackage{amsmath}
\usepackage{amsthm}
\usepackage[warn]{mathtext}
\usepackage{braket}
\usepackage{bm}
\usepackage{enumitem}
\usepackage{tabularx}
\usepackage{indentfirst}
\usepackage{wrapfig}
\usepackage{topcapt}
\usepackage{comment}
\usepackage{subfigure}
\usepackage{color,xcolor}
\definecolor{dark-blue}{rgb}{0,0,0.875}
\definecolor{dark-green}{rgb}{0,0.625,0}
\definecolor{dark-red}{rgb}{0.875,0,0}
\usepackage[colorlinks=true,linkcolor=dark-red,filecolor=dark-blue,
citecolor=dark-green,urlcolor=dark-blue,breaklinks=true]{hyperref}
\begin{document}
\title{Multi-party quantum summation based on quantum teleportation}
\author{Cai Zhang $^{1,2,*}$, Mohsen Razavi $^{2,*}$, Zhiwei Sun $^{3,4,*}$, Qiong Huang $^{1}$ and Haozhen Situ $^{1}$}
\affiliation{%
$^{1}$ College of Mathematics and Informatics, South China Agricultural University, Guangzhou, 510642, China \\
$^{2}$ School of Electronic and Electrical Engineering, University of Leeds, Leeds, LS2 9JT, UK \\
$^{3}$ School of Artificial Intelligence, Shenzhen PolyTechnic, Shenzhen, 518055, China \\
$^{4}$ Center for Quantum Computing, Peng Cheng Laboratory, Shenzhen 513055, China 
}


\begin{abstract}
We present a secure multi-party quantum summation protocol based on quantum teleportation,
	in which a malicious, but non-collusive, third party (TP) helps compute the summation. In our protocol, TP is in charge of entanglement distribution and Bell states are shared between participants. Users encode the qubits in their hand according to their private bits and perform Bell-state measurements. After obtaining participants' measurement results, TP can figure out the summation. The participants do not need to send their encoded states to others, and the protocol is therefore congenitally free from Trojan horse attacks. In addition, our protocol can be made secure against loss errors, because the entanglement distribution occurs only once at the beginning of our protocol. We show that our protocol is secure against attacks by the participants as well as the outsiders.
\end{abstract}
\maketitle
\section{Introduction}
Secure multi-party computation, as a subfield in cryptography, has been gaining attention in recent years \cite{halevi2016secure,baum2016better,ben2016optimizing,keller2018efficient}. It was first introduced by Yao \cite{yao1982protocols} and later extended by Goldreich et al. \cite{GMW87}. Secure multi-party computation has also been studied in quantum settings \cite{lo97,crepeau02secure,chau00quantum,ben06secure,smith10multi}.  Lo \cite{lo97} pointed out the insecurity of quantum computation without a third party in a two-party scenario. 
Chau \cite{chau00quantum} employed quantum resources to speed up classical multi-party computation.
Ben-Or et al. \cite{ben06secure} investigated distributed quantum computation. They showed how many players must be honest in order to make any multi-party quantum computation secure.
Smith \cite{smith10multi} proved that any multi-party quantum computation can be secure as long as the number of dishonest players is less than $n/6$, when $n$, the number of players, is larger than $6$.

Secure multi-party quantum summation \cite{H02,HN03,HKW03,DCWZ07,chen10efficient}, which helps the construction of complex multi-party computation, is a fundamental primitive of secure multi-party quantum computation.  In quantum summation protocols, the privacy of participants' inputs is preserved and the correctness of the summation is guaranteed by quantum properties. Quantum summation has also potential applications in quantum voting \cite{hillery06towards,li08quantum,wang16self,xue17simple,bao17quantum} and quantum private equality comparison \cite{sun15quantum,hung17multiparty,he17quantum}. Designing quantum summation protocols that can be implemented with current or near future quantum technologies is therefore of interest, as we pursue in this paper.

In the past few years, various quantum summation protocols have been proposed by employing a variety of quantum resources. Zhang et al. \cite{zhang2014high} presented a quantum summation protocol with single photons encoded in both polarization and spatial-mode degrees of freedom in 2014, in which unitary operations are utilized to encode the private bits on the travelling single photons. Such single photons must somehow be handed over/transmitted to the next user so that the collective sum of all private bits can be calculated. Most other protocols rely on sharing a multipartite entangled state among players. For instance, in 2015, a quantum summation protocol without a trusted third party was constructed \cite{zhang2015three}. However, the number of participants was limited to three due to the requirement of the so-called genuinely maximally entangled six-qubit states. In 2016, Shi et al. \cite{shi2016secure} used quantum Fourier transform, controlled NOT (CNOT) gates and oracle operators to propose protocols for summation and multiplication. Later, they proposed a common quantum solution to a class of two-party private summation problems \cite{shi2017quantum}. In 2017, a multi-party quantum summation without a trusted third party was investigated by first generating a multipartite entangled state by one player and then sharing it with other users \cite{zhang2017multi}. In the same year, Liu et al. \cite{liu2017novel} adopted Bell states to construct multipartite entangled states that were used to carry participants' inputs, where the quantum communication in their protocol is two-way. This means that special care with regard to Trojan horse attacks \cite{Deng05,Gisin06,Li06Deng} should be provided to participants. Unlike their protocols, participants in our protocol do not need to send the encoded states back to others, thus our protocol is naturally free from Trojan horse attacks and no protection against such attacks are needed. In 2018, Yang et al. \cite{yang2018secure} provided a quantum solution to secure summation depending on $n$-partite multi-dimensional entangled states.

One common feature in all hitherto proposed quantum summation protocols is their dependence on a reliable means for quantum state transfer. In the case of protocols that rely on sharing multipartite entangled states \cite{shi2016secure, shi2017quantum, zhang2017multi, liu2017novel, yang2018secure}, such a state is often generated by one player and then its different components are sent to other players. If any of these components does not reach its respective destination, then the whole procedure must be repeated. In such a case, relying on photons travelling through lossy channels does not seem to be an efficient option. Moreover, it could open us to new security threats that an eavesdropper can exploit by hiding behind the channel loss. Even for the case of the protocol in Ref.~\cite{zhang2014high}, the loss of the single photon in any leg of the system requires repeating the whole procedure. In addition, an eavesdropper can send a photon of her choice to a user and measure it after the user has applied his encoding to find out about the user's private bit. Most these protocols fail to work unless a reliable quantum state transfer (RQST) service is available to them. This is a kind of service that one may expect to have once we have a fully functional quantum network.

There are two well-known approaches to RQST. In one scenario, one distributes entangled states between the two end users of a quantum communication system, and then use teleportation to transfer an unknown quantum state from one place to another. In the second approach, one has to use perhaps complex quantum error correction codes to compensate for the erasure errors caused by photon loss as well as operational errors caused by system components. In both cases, we need quantum memories in our setup to store quantum states and to execute certain quantum processing tasks such as entanglement distillation or quantum error correction. This requirement of the system has thus far been neglected in the design of quantum summation protocols.

In this paper, we take advantage of the idea of quantum teleportation \cite{bennett93teleporting} to devise our protocol. In order to get a better insight into the practicality of a quantum summation protocol, in this work, we account for the bipartite entangled states that one would need to distribute if teleportation is used for the RQST part of the protocol. We discover that in fact such Bell states are sufficient to devise a secure quantum summation protocol without requiring the distribution of additional multipartite entangled states. Moreover, by not revealing the information about which Bell state is shared between two players, we, in effect, can protect ourselves against attacks by malicious participants. In our protocol, similar to Ref.~\cite{zhang2014high}, participants' private bits are encoded into single-qubit unitary operations. Encoded states are then effectively teleported to the next user by performing local Bell-state measurements (BSMs). This makes our protocol congenitally free from Trojan horse attacks. In our protocol, the required Bell states are shared by a third party (TP), who can be malicious but does not collide with other players. In any case, our protocol does not rely on multipartite entanglement or high-dimensional states, which makes its implementation much more feasible. 

Table~\ref{tab.comp} summarizes the required resources for various protocols as compared to ours. In particular, we have compared these protocols in terms of their efficiency, defined as the number of qubits (quantum memories) they need in order to find the sum of $n$ private bits, when one accounts for a minimum of two quantum memories needed for teleportation. The assumption here is that maximally entangled states are shared among users, but we do not account for additional memories that may be needed for entanglement distillation or for possible repeater nodes. It is clear from this table that our protocol not only is more efficient than other protocols in the table but also only relies on bipartite entanglement rather than multipartitite states.

\begin{table*}[t]
\footnotesize
\caption{ A comparison between different quantum summation (QS) protocols in terms of their required resources and operations, as well as their efficiency.}
\label{tab.comp}
\tabcolsep 12pt 
\begin{tabular*}{0.9\textwidth}{m{8em} m{1.9em} m{13em}  m{6cm} }
\toprule
QS protocols & Efficiency   & Quantum resource & Quantum Operations \\\hline

Shi et al.'s \cite{shi2016secure} & $\frac{1}{3n-2}$  & $(n+1)$-partite entangled state &
Quantum Fourier operator,  CNOT operator, and oracle operator  \\ \hline

Zhang et al.'s \cite{zhang2017multi}  & $\frac{1}{3n-2}$ &  $n$-partite entangled state &
  CNOT operator and Hadamard operator  \\ \hline
  
Liu et al.'s \cite{liu2017novel}  & $\frac{1}{3n-2}$ or $\frac{1}{3n+1}$ & $n$-partite entangled state or $(n+1)$-partite entangled state& Pauli operators and Hadamard operators   \\ \hline

Yang et al.'s \cite{yang2018secure} & $\frac{1}{3n-2}$& $n$-partite entangled state & Quantum Fourier operator and Pauli operators   \\ \hline

 This work & $\frac{1}{2n+3}$ & Bell states & Pauli operators and Bell measurement   \\
 \hline
\end{tabular*}
\end{table*}


The rest of this paper is organized as follows. In Sec. \ref{sec2two}, we illustrate our idea to design a secure multi-party quantum summation protocol and provide an example of a two-party scenario. In Sec. \ref{secmul}, we describe our multi-party quantum summation protocol in detail, followed by its correctness and security analysis in Sec. \ref{analysis}. Practical considerations of our protocol will be discussed in Sec. \ref{sec.practical}, and conclusions are given in Sec.~\ref{secdiccon}.

\section{Key Idea of Our Protocol} \label{sec2two}
In this section, we work out our proposed quantum summation protocol for the particular case of two participants and a malicious but non-collusive third party (TP). TP has to calculate the modulo 2 sum of the participants' secret bit by satisfying the following requirements:
\begin{enumerate}
	\item Correctness: The result of summation in modulo two of all participants' private input bits is correct.
	\item Security: An eavesdropping outsider cannot learn any information about participants' private input bits without being detected.
	\item Privacy: TP cannot learn about participants' private input.
\end{enumerate}
Note that although TP cannot obtain two participants' private bits in the two-party scenario, each participant can find out the private bit of the other participant once the sum is known. Nevertheless, this is a simple example by which we can explain our protocol. In Sec. \ref{secmul}, we generalize this idea to multiple participants scenario, where the privacy requirement will be extended to include most participants as well as TP.

\begin{figure}
	\centering
	\includegraphics[width=1\textwidth]{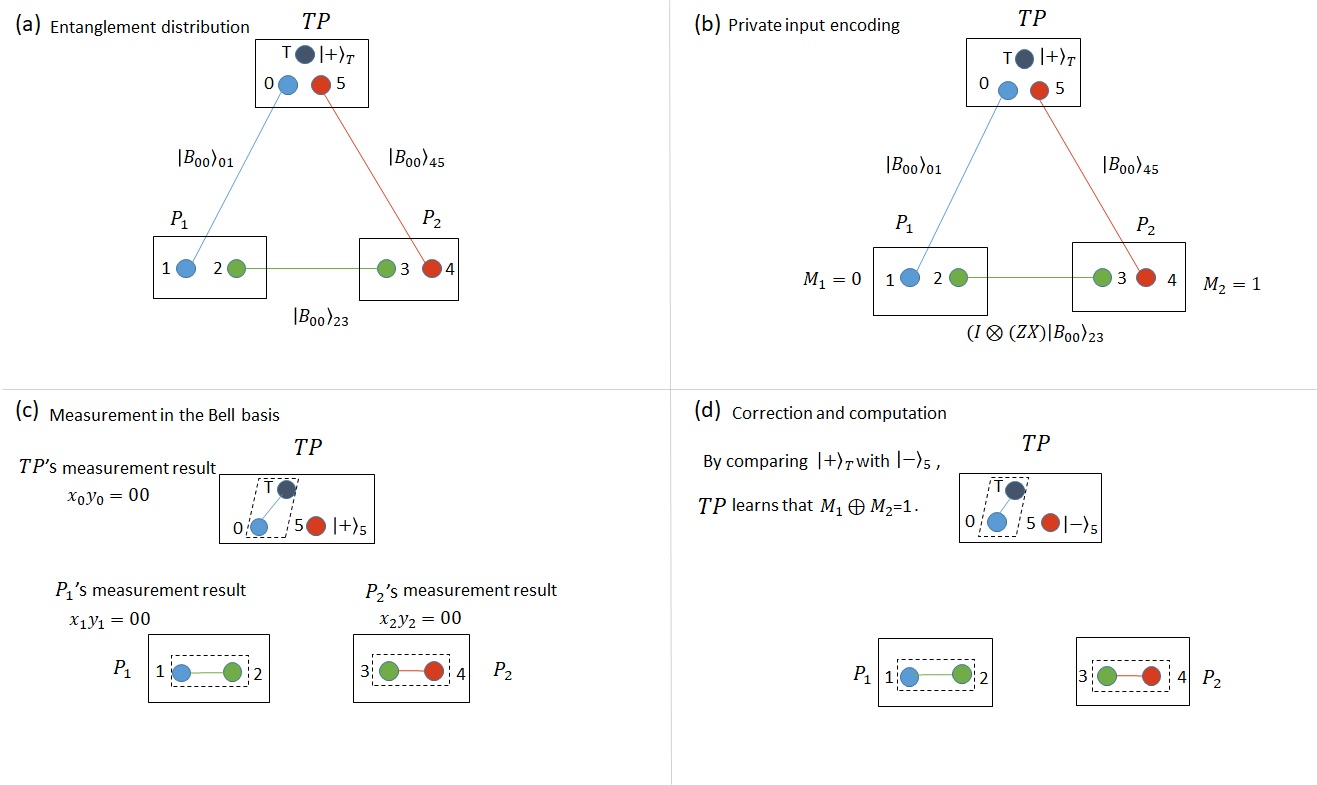}
	\centering
	\caption{A simple example of our protocol in the two-party scenario. (a) Step 1: TP shares entangled states among users to create a chain of entangled links back to herself. In this example, we assume state $|B_{00}\rangle$ is shared over all links. In general, different Bell states can be shared over different links, and only TP knows which state has been shared. (b) Step 2: Users with private bit 1 apply operator $U$ to thier first qubit. Here, only $P_2$ must do this. (c) Step 3: All players perform a BSM on their two qubits and let TP know of the results. In our example, we have assumed $|B_{00}\rangle$ has been obtained in all cases. (d) Step 4: TP measures qubit 5 in the same basis as her originally chosen basis for qubit $T$. By comparing the result with the original state of $T$, TP can calcualte $M_1\oplus M_2$. }\label{fig.pro}
\end{figure}

Our protocol relies on sharing a chain of Bell states among participants and teleporting an unknown state by TP to itself via this chain; see Fig.~\ref{fig.pro}. Along the way participants can affect the linked states by applying local operations on their share of entangled states. TP can calculate the sum by comparing the teleported state with the original state she has generated.

Before describing the protocol, let us first review the teleportation protocol and introduce the notation used in the paper. In general, Bell states are of the following form 
\begin{equation}\label{eq.belsta}
|B_{xy}\rangle=\frac{1}{\sqrt{2}}(|0,x\rangle+(-1)^{y}|1,x\oplus 1\rangle),
\end{equation}
where $x,y\in \{0,1\}$ and $ \oplus $ represents addition modulo 2. The relationship between Bell states and classical bits can be defined as
\begin{equation}\label{eq.bellclass}
|B_{xy}\rangle \leftrightarrow xy, x,y \in \{0,1\}.
\end{equation}
For any qubit $ |\varphi\rangle$ and any single-qubit unitary operation $U$, a general teleportation equation, based on an initial Bell state $|B_{ab}\rangle, a,b \in \{0,1\}$, shared between the two users, can be written as 
\begin{equation}\label{eq.tel}
|\varphi\rangle_{1}\otimes(I \otimes U)|B_{ab}\rangle_{2,3} = 
\frac{1}{2}\sum_{x \in \{0,1\}}\sum_{y \in \{0,1\}}(-1)^{b\cdot x}|B_{xy}\rangle_{1,2}\otimes UZ^{y\oplus b}X^{x\oplus a}|\varphi\rangle_{3},
\end{equation}
where $X=(|0\rangle\langle 1|+|1\rangle\langle 0|) $,
$Z=(|0\rangle\langle 0|-|1\rangle\langle 1|)$ and the subscripts denote different systems. 

In this work, we are particularly interested in the unitary operation $U=ZX$, for which we have:
\begin{equation}\label{eq.swa}
\begin{array}{ll}
UZ^{b}X^{a} &= ZX Z^{b}X^{a} = (-1)^{b}Z^{b}ZXX^{a} = (-1)^{b}Z^{b}ZX^{a}X\\
&=(-1)^{b}\cdot (-1)^{a} Z^{b}X^{a}ZX = (-1)^{a\oplus b} Z^{b}X^{a}ZX \\
&=(-1)^{a\oplus b} Z^{b}X^{a}U,
\end{array}
\end{equation}
where $a,b\in \{0,1\}$. Additionally, the following equations 
\begin{eqnarray}\label{eq.four}
U|0\rangle &=& ZX|0\rangle=-|1\rangle,\\
U|1\rangle &=& ZX|1\rangle= |0\rangle,\\
U|+\rangle &=& ZX|+\rangle= |-\rangle,\\ \label{eq.four.last}
U|-\rangle &=& ZX|-\rangle=-|+\rangle,
\end{eqnarray}
hold, where $|+\rangle=\frac{1}{\sqrt{2}}(|0\rangle+|1\rangle)$ and 
$|-\rangle=\frac{1}{\sqrt{2}}(|0\rangle-|1\rangle)$. Note that 
both computational basis $ \{|0\rangle,|1\rangle\} $ and diagonal basis $\{|+\rangle,|-\rangle\} $ are closed under $ U $. Ignoring the phase, $U$ swaps $|0\rangle$ and $|1\rangle$ ($|+\rangle$ and $|-\rangle$).  We use $U=ZX$ from now on and it will be applied on one of the two components of a Bell state if the participants' private bit is $1$.

Now, let us describe a simple version of our protocol that, for now, does not fulfill the security requirement; see Figure \ref{fig.pro}. Suppose each participant has two quantum memories. Then, we implement the following steps:
\begin{enumerate}[leftmargin=1.2cm,labelsep=2mm,label=(Step {\arabic*})]
	\item \textit{Entanglement distribution}. TP distributes Bell states, each of which is randomly selected from the Bell basis, among participants and generates a state $|\varphi\rangle_{T}$ chosen randomly from the set $\{|0\rangle,|1\rangle,|+\rangle,|-\rangle\}$. The state $|\varphi\rangle_{T}$ is stored in quantum memory $T$.
	\item \textit{Private inputs encoding}.  $P_{1}$ ($P_{2}$) applies $U=ZX$ on quantum memory 1 (quantum memory 3) if her private bit is $1$. Otherwise, she does nothing.
	\item \textit{Bell-state measurement.} TP measures quantum memories $T$ and $0$ in the Bell basis. Similarly, $P_{1}$ ($P_{2}$) measures quantum memories $1$ and $2$ ($3$ and $4$) in the Bell basis. $P_{1}$ and $P_{2}$ will announce their measurement results to TP.
	\item \textit{Correction and computation.} After necessary corrections on quantum memory $5$ depending on all the measurement results and the original Bell states, TP measures quantum memory 5 in the same basis as that of the original state of quantum memory T. If the state of quantum memory 5 is the same as the original state of quantum memory T, TP concludes that the sum is $0$, otherwise, the sum is $1$.
\end{enumerate}

Let us work out a simple example to show how the protocol works. In Figure \ref{fig.pro},
\begin{enumerate}[leftmargin=1.2cm,labelsep=2mm,label=(Step {\arabic*})]
	\item \textit{Entanglement distribution}. Suppose the initial state among TP, $P_{1}$ and $P_{2}$ is given by
	\begin{equation}\label{eq.corini.simple} 
	|\zeta_{j}^{0}\rangle=|+\rangle_{T}\otimes |B_{00}\rangle_{01}\otimes  |B_{00}\rangle_{23}\otimes |B_{00}\rangle_{45}.
	\end{equation}
	\item \textit{Private input encoding}. Suppose $P_{1}$'s ($P_{2}$'s) private bit is $0$ ($1$),	 
	$P_{1}$ then does nothing on quantum memory 1, but $P_{2}$ applies $U=ZX$ on quantum memory 3. According to Eqs. (\ref{eq.tel}-\ref{eq.four.last}), the state becomes
	\begin{align}\label{eq.twoparty}
	|\zeta_{j}^{1}\rangle=&|+\rangle_{T}\otimes 
	(I\otimes I)|B_{00}\rangle_{01}\otimes  
	(I\otimes (ZX))|B_{00}\rangle_{23}\otimes 
	|B_{00}\rangle_{45} \nonumber \\ 
	= &\frac{1}{8} \sum_{x_{0}\in \{0,1\}} \sum_{y_{0}\in \{0,1\}} \sum_{x_{1}\in \{0,1\}} \sum_{y_{1}\in \{0,1\}} \sum_{x_{2}\in \{0,1\}} \sum_{y_{2}\in \{0,1\}} \nonumber \\ 
	&|B_{x_{0}y_{0}}\rangle_{T0}|B_{x_{1}y_{1}}\rangle_{12}|B_{x_{2}y_{2}}\rangle_{34} Z^{y_{1}\oplus y_{2}\oplus y_{3}}X^{x_{1}\oplus x_{2}\oplus x_{3}}|-\rangle_{5},
	\end{align}
	where a global phase in the state of quantum memory 5 is ignored. 
	\item \label{enu.exm}
	\textit{Bell-state measurement.} Suppose all the measurement results are  
	$x_{0}y_{0}=x_{1}y_{1}=x_{2}y_{2}=00$, and they are announced to TP. Then, effectively, the state of $T$ is teleported to qubit 1, and then teleported to qubit to 3, at which point it is flipped by the $U$ operation, and teleported back to TP. 
	\item \textit{Correction and computation.} In this particular case, there is no correction needed by TP. TP measures quantum memory 5 in the basis $\{|+\rangle,|-\rangle\}$, and finds that the state of quantum memory 5 is different from the original state of quantum memory $T$. TP concludes that the sum is $1$.
\end{enumerate}
In \ref{enu.exm} of the above example, if not all the measurement result are $00$, TP can correct the state of quantum memory 5 by performing quantum operations on it using Eqs. (\ref{eq.tel}-\ref{eq.swa}) before she measures quantum memory 5.

In a full protocol, we need to include steps that alert us to possible attacks. We consider two kinds of attacks in our protocol: those by outsides and those by malicious participants. We employ extra Bell states to detect these attacks and meet the security requirements. By measuring each component of a Bell state in the same basis (all in the computational basis or all in the diagonal basis) and comparing the measurement results, these attacks can be detected. The details of the detection process can be found in Sec.~\ref{secmul}.

\section{Multi-Party Quantum Summation} \label{secmul}
We assume that the classical channels are authenticated and quantum channels are noiseless. The third party, TP, who conducts the summation is assumed to be malicious but non-collusive. That is to say, TP can do whatever she would like within boundaries of quantum mechanics except collision with dishonest participants. The summation can be revealed in public. For simplicity, we denote TP as $ P_{0} $ in the rest of the paper.

Suppose that the $ q $-th participant ($q=1,2,\ldots,n$; $n>2$) has a private bit string 
$ M_{q} $. $ P_{0} $ computes the summation
$ \oplus\sum_{q=1}^{n} M_{q}$, where  $\oplus\sum$  denotes pointwise addition in modulo $2$, and
\begin{equation}\label{eq.mul2}
\begin{array}{rcl}
M_{1} &=& (m_{11},m_{12},\ldots,m_{1L}), \\
M_{2} &=& (m_{21},m_{22},\ldots,m_{2L}), \\
&\ldots&, \\
M_{n} &=& (m_{n1},m_{n2},\ldots,m_{nL}),\\
\oplus\sum_{q=1}^{n} M_{q}&=&(\sum_{i=1}^{n}m_{i1},\sum_{i=1}^{n}m_{i2},\ldots,\sum_{i=1}^{n}m_{iL}),
\end{array}
\end{equation}
where $L$ is the length of each private bit string.

\begin{figure}
	\centering
	\includegraphics[width=1\textwidth]{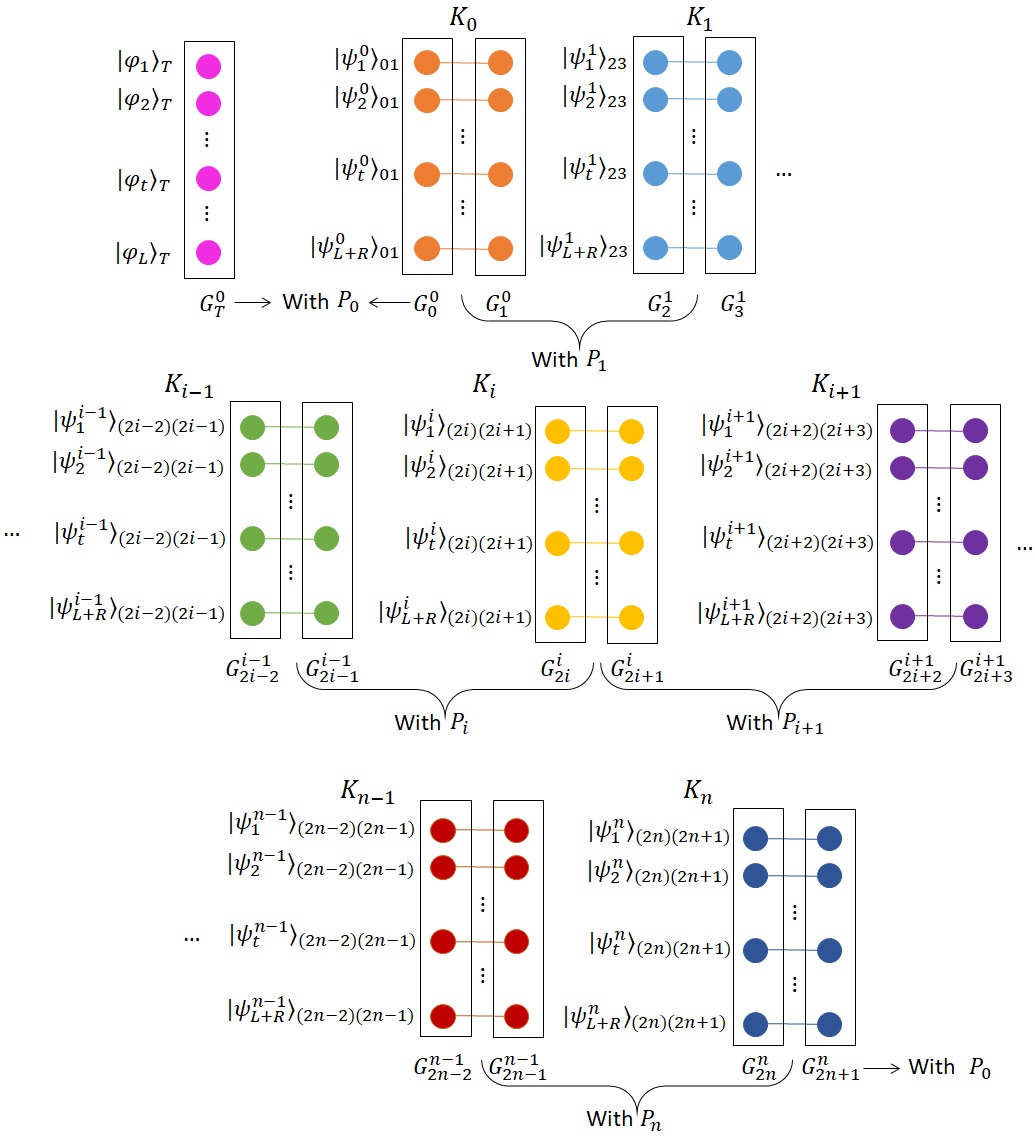}
	\centering
	\caption{Entanglement distribution by $P_{0}$. Each player has a qubit which is entangled with another qubit held by the next user in the chain. At the start of the protocol, TP shares $L+R$ Bell states over each link, where $R$ of which (randomly chosen) is used for detecting malicious activities.
	}\label{fig.step1}
\end{figure}
Our $n$-party ($n>2$) summation protocol shall meet the following requirements:
\begin{enumerate}[leftmargin=1.2cm, labelsep=2mm]
	\item Correctness: The result of pointwise summation in modulo two of all participants' private input bits is correct.
	\item Security: An outside eavesdropper cannot learn any information about participants' private input bits without being detected.
	\item Privacy: No participant can learn  about other participants' private input bits without being detected, except in the obvious case of $n-1$ players collaborating to learn the remaining user's private bits.
\end{enumerate}

Our full protocol is described in the following.
\begin{enumerate}[leftmargin=1.2cm,labelsep=2mm, label=(Step {\arabic*})]
	\item \label{enu.fullStep1}
	\textit{Entanglement distribution.}
	$P_{0}$ uses a certain entanglement distribution protocol \cite{sangouard2011quantum,razavi2007nonadiabatic,amirloo2010quantum,lo2013quantum,bruschi2014repeat} to distribute $(n+1)(L+R)$ ordered Bell states,
	$K_{i}=(|\psi_{1}^{i}\rangle_{(2i)(2i+1)}|\psi_{2}^{i}\rangle_{(2i)(2i+1)}\ldots$  $|\psi_{L+R}^{i}\rangle_{(2i)(2i+1)})$ 
	($i=0,1,\ldots,n$), where $|\psi_{1}^{i}\rangle_{(2i)(2i+1)}$ is chosen from the set $\{|B_{xy}\rangle|x,y\in \{0,1\}\}$, to $n$ participants such that these states form a chain.  Specifically, for $K_{i}$, all first (second) components of Bell states are stored in quantum memory $G_{2i}^{i}$ ($G_{2i+1}^{i}$). As shown in Figure \ref{fig.step1}, banks of quantum memories $G_{2i-1}^{i-1}$ and $G_{2i}^{i}$ belong to $P_{i}$ ($i=1,2,\ldots,n$) and quantum memories $G_{0}^{0}$ and $G_{2n}^{2n+1}$ are held by $P_{0}$. 
	$P_{0}$ also generates $L$  ordered states, $A_{T}=(|\varphi_{1}\rangle_{T},|\varphi_{2}\rangle_{T},\ldots,$
	$|\varphi_{L}\rangle_{T})$, where
	$ |\varphi_{i}\rangle_{T} $ ($i=1,2,\ldots,L$) is randomly chosen from the set $\{|0\rangle, |1\rangle,|+\rangle,|-\rangle\}$. These states remain in $P_{0}$'s quantum memory $G^{0}_{T}$. Note that all the initial states are only known to $ P_{0} $.
	\item \label{enu.fullStep2}
	\textit{Security detection.}
	Participants detect if genuine Bell states are shared among them in an honest way. 
	\begin{enumerate}[label=(Step {2.\arabic*})]
		\item To examine the genuinity of the Bell states shared between $ P_{0} $ and $ P_{1} $,	$P_{1}$ first randomly chooses $R$ Bell states shared between quantum memory $G_{0}^{0}$ and quantum memory $G_{1}^{0}$ and asks $P_{0}$ to announce the corresponding initial states. $P_{1}$ then measures each corresponding component in $G_{1}^{0}$ randomly in the computational basis 
		$\{|0\rangle, |1\rangle\}$ or in the diagonal basis $ \{|+\rangle,|-\rangle\}$, and keeps the measurement results to herself. Subsequently, $P_{1}$ asks $P_{0}$ to measure the corresponding components in the same basis as $ P_{1} $ does and publicize the measurement results. According to the property of Bell states, $P_{1}$ checks if these measurement results are correlated with each other. If the error rate exceeds a certain threshold, the protocol will be aborted and repeated from \ref{enu.fullStep1}. Otherwise, the protocol will continue. 
		\item To check the genuinity of the Bell states shared between $ P_{0} $ and $ P_{n} $,
		$P_{n}$ also uses $R$ Bell states to complete this detection utilizing the similar method as that used by $ P_{1} $. If the error rate exceeds the threshold, the protocol will be aborted and repeated from \ref{enu.fullStep1}. Otherwise, the protocol will continue. 
		\item To check the genuinity of the Bell states shared between $ P_{i} $ and $ P_{i+1}$ 
		($i=1,2,\ldots,n-1$),
		$P_{i}$ randomly selects $R/2$ Bell states shared between $G_{2i}^{i}$ and $G_{2i+1}^{i}$ and asks $P_{0}$ to announce the corresponding initial states. Later, $P_{i}$ measures each corresponding component in $G_{2i}^{i}$ randomly in the computational basis 
		or in the diagonal basis, announcing the measurement results. Next, $P_{i+1}$ measures each component in $ G_{2i+1}^{i} $ entangled with the one in $ P_{i}$'s hands in the same basis, publicizing the measurement results. $P_{i}$ and $P_{i+1}$ can finally check if these measurement results are correlated according to the initial states and the property of Bell states. The same procedure will be used by $P_{i+1}$ with $R/2$ Bell states of his choice and randomly selected measurement bases. If the error rate in either case exceeds the threshold, the protocol will be aborted and repeated from \ref{enu.fullStep1}. Otherwise, they ensure that the states shared between them are genuine Bell states and distributed in an honest way, and the protocol will continue. 
	\end{enumerate}
	\item \label{enu.encoding}
	\textit{Private input encoding.}
	$P_{0}$ removes $R$ states used for detection from quantum memory
	$ G_{0}^{0} $ ($ G_{2n+1}^{n} $), leaving $L$ ordered states, denoted by $V^0_0$ ($V^{n}_{2n+1}$), in it.
	$ P_{i} $ ($ i=1,2,\ldots,n$) also
	removes $R$ states used for checking from quantum memory
	$  G_{2i-1}^{i-1} $ ($G_{2i}^{i} $), resulting in $L$ ordered states, denoted by $V_{2i-1}^{i-1}$ ($V_{2i}^{i}$), in it. Note that quantum memories
	$ G_{2i}^{i} $ and $ G_{2i+1}^{i} $ ($ i=0,1,\ldots,n$) now share $L$ ordered Bell states, which form $L$ chains of Bell states among all participants (inlucding $P_{0}$). 
	Namely, the $j$-th ($ j=1,2,\ldots, L$) state of $ V_{2i}^{i}$ in $ G_{2i}^{i}$ and the 
	$ j $-th one of $V_{2i+1}^{i}$ in $ G_{2i+1}^{i}$ form a Bell state. Afterwards,
	$ P_{i} $ ($ i=1,2,\ldots,n $) performs 
	$ U_{i}^{m_{i1}} \otimes  U_{i}^{m_{i2}} \otimes \ldots \otimes U_{i}^{m_{iL}} $ on the ordered sequence 
	$ V_{2i-1}^{i-1} $, where 
	$ U_{i}=U=ZX$ and $(m_{i1},m_{i2},\ldots,m_{iL})$ is $P_{i}$'s private bit string.
	\item \label{enu.fullStep4}
	\textit{Bell-state measurement.}
	$P_{0}$  measures the $j$-th ($j=1, 2,\ldots, L$) state of
	$ V_{0}^{0} $ and the $j$-th one in quantum memory 
	$ G^{0}_{T} $ in the Bell basis, obtaining measurement results
	($ x_{01}y_{01}, x_{02}y_{02}, \ldots, x_{0L}y_{0L} $) in accordance with Eq. (\ref{eq.bellclass}). Similarly,
	$P_{i}$ ($ i=1,2,\ldots,n$) measures the $j$-th state of $ V_{2i-1}^{i-1} $ and the 
	$j$-th one of $ V_{2i}^{i}$ in the Bell basis, attaining measurement results
	($x_{i1}y_{i1}, x_{i2}y_{i2}, \ldots, x_{iL}y_{iL}$). Finally, They announce the measurement results to $P_{0}$.
	\item 
	\textit{Correction and computation.}
	Based on all the measurement results and the knowledge of original Bell states (only known to $P_{0}$), $P_{0}$ performs correcting operations on the $j$-th ($j=1, 2, \ldots, L$) state of $ V_{2n+1}^{n} $. Next, $P_{0}$ measures these resulted states in the same basis as the original states in quantum memory $ G^{0}_{T} $, gaining the measurement results ($t_{1}, t_{2}, \ldots, t_{L}$).
	With these measurement results, $P_{0}$ compares the $j$-th  state of $V_{2n+1}^{n}$ with the $j$-th original state in quantum memory $ G^{0}_{T} $. If these two states are the same (different), $P_{0}$ knows that the $j$-th bit of the sum is $0$ ($1$). At last, 
	$P_{0}$ can achieve the sum modulo $2$ of participants' private bit strings, and the privacy of these private strings is preserved.
\end{enumerate}

Note that, if the summation is only intended for a certain participant, say $ P_{i} $, she can be selected as the one who distributes Bell states like TP. The process is analogous to that with TP if $ P_{i} $ is also assumed to be malicious, but non-collusive.

\section{Analysis of the Multi-party Quantum Summation}\label{analysis}
In this section, we study the security of our protocol. It can be verified that the protocol would provide us with the correct sum if all parties follow the protocol. A detailed derivation of the correctness is given in Appendix~\ref{app.analysis}. In terms of security, we have to show that our protocol is secure against both outsider and participant attacks, and it fulfills the security and privacy requirements mentioned in Sec. \ref{secmul}. In our case, an outsider can potentially influence our protocol via the initial entanglement distribution. We show here how by using extra Bell states we can verify if the distributed states are genuinely Bell states. There also exist Trojan horse attacks \cite{Deng05,Gisin06,Li06Deng}, such as the delay-photon Trojan horse attack and the invisible photon eavesdropping Trojan horse attack if quantum states are encoded and relayed in quantum communications protocols. Since our protocol uses Bell states to compute the summation and no encoded states are needed to be relayed, our protocol is secure against these attacks. We therefore, here, focus on the case of an attack by the TP, or possibly an outsider, and leave the details of the security against other malicious participants to Appendix~\ref{app.analysis}.



\begin{figure}
	\centering
	\includegraphics[width=0.6\textwidth]{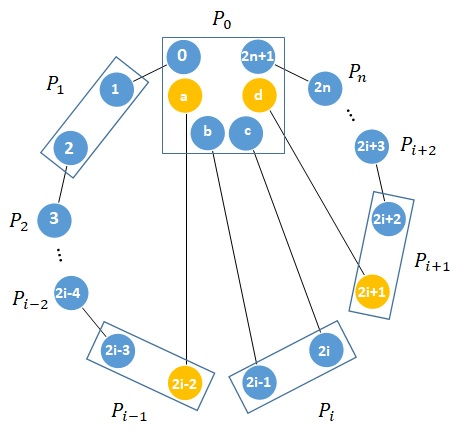}
	\centering
	\caption{Entanglement swapping attack by $P_{0}$ through sharing entangled states in a dishonest way.}\label{fig.attack.p0}
\end{figure}

\textit{Attacks from $P_{0}$.} We here consider the attacks from $P_{0}$ who cannot collude with any other participants. For simplicity, we suppose that $ P_{0} $ wants to obtain one bit of $P_{i}$'s ($i\ne 1, n$) private bit string and consider the chain related to this bit. In order to learn about this bit of $P_{i}$, $P_{0}$ has to find out if $P_{i}$ performs quantum operation $U=ZX$ on her memory. $P_{0}$ can therefore launch entanglement swapping attack on this chain, as shown in Fig.~\ref{fig.attack.p0}. 

Suppose, in Fig.~\ref{fig.attack.p0}, the states of quantum memories $b$ and ($2i-1$) and quantum memories $c$ and ($2i$) distributed by $P_{0}$ are 
$|B_{00}\rangle_{b(2i-1)}$ and $|B_{00}\rangle_{c(2i)}$, respectively. $P_{i}$ will apply $U=ZX$ on quantum memory ($2i-1$) if her secret bit is $1$, otherwise she will do nothing. $P_i$ then measures quantum memories ($2i-1$) and ($2i$) in the Bell basis and announces her measurement result $x_{i}y_{i}$ to $P_{0}$ as described in \ref{enu.fullStep4} in the proposed protocol. After that, $P_{0}$ can measure quantum memories $b$ and $c$ as well and obtain the measurement result $x_{c}y_{c}$. Because the original states of quantum memories $b$ and ($2i-1$) and quantum memories $c$ and ($2i$) are the same, if $x_{i}y_{i}$ and $x_{c}y_{c}$ are the same, $P_{0}$ knows that $P_{i}$ has not performed $U$ on quantum memory ($2i-1$) and learns about $P_{i}$'s private bit being $0$, according to the entanglement swapping property. Otherwise, $P_{0}$ concludes that $P_{i}$'s private bit is $1$. However, this attack will be detected in \ref{enu.fullStep2} where the genuinity of Bell states shared between $P_{i}$ and $P_{i+1}$ (between $P_{i-1}$ and $P_{i}$) is checked.

To show this note that Bell states can be rewritten in linear and diagonal bases as follows
\begin{eqnarray}\label{eq.beldif}
|B_{00}\rangle=\frac{1}{\sqrt{2}}(|00\rangle+|11\rangle)=\frac{1}{\sqrt{2}}(|++\rangle+|--\rangle),\\
|B_{01}\rangle=\frac{1}{\sqrt{2}}(|00\rangle-|11\rangle)=\frac{1}{\sqrt{2}}(|+-\rangle+|-+\rangle),\\
|B_{10}\rangle=\frac{1}{\sqrt{2}}(|01\rangle+|10\rangle)=\frac{1}{\sqrt{2}}(|++\rangle-|--\rangle),\\
|B_{11}\rangle=\frac{1}{\sqrt{2}}(|01\rangle-|10\rangle)=\frac{1}{\sqrt{2}}(|-+\rangle-|+-\rangle).
\end{eqnarray}
If $P_{i}$ and $P_{i+1}$ shared a known Bell state, and each one measures one component of the Bell state in the same basis (in the computational basis or in the diagonal basis), they will obtain a certain relationship between their measurement results.  For a fake Bell state (the state of quantum memories ($2i-1$) and ($2i-2$) is not a Bell state, we call it a fake Bell state) used for detection, $ P_{0} $ is able to pass the detection with probability of $\frac{1}{2}$. $P_{0}$ may distribute only one fake Bell state between $P_{i}$ and $P_{i+1}$ and another fake Bell state between $P_{i-1}$ and $P_{i}$ such that these two states are in the same chain to obtain $P_{i}$'s private bit. At the same time, $P_{0}$ can get the maximum probability of passing the detection. In this case, these two states should not be chosen for detection. The probability of escaping the detection is
$L^{2}/(L+R)^2$. For $i=1$ or $i=n$, this probability becomes $L/(L+R)$. These two probabilities of $ P_{0} $ passing the detection and obtaining one bit of one participant will approach  $ 0 $ if $ R $ is large enough. As a result, $ P_{0} $ fails to steal participants' private input bits.

\section{Practical Considerations} \label{sec.practical}
In this section, we discuss some practical aspects of our protocol in the light of new developments in the field. In general, secure multi-party quantum computation requires an infrastructure for reliable quantum communications as provided by quantum repeaters and quantum networks. Our protocol is not an exception, but given that some of the required resources for our protocol, as listed in Table~\ref{tab.comp}, are easier to achieve, we can envisage a small-scale demonstration of this protocol in the near future. Multicore optical fibres \cite{Bacco17Ding,Eriksson17Hirano} can be used to fish this task.

One of the key requirements in our scheme is to distribute Bell states between two parties. A full implementation of this aspect over any arbitrary distance is only possible with fully functional quantum repeaters. This may not be possible in the near future. But, a small-scale quantum network with nodes within tens of km from each other is within reach. In fact, there are activities in Netherlands, for instance, to implement a four node quantum network within the country. Such a network can then be used for an initial demonstration of protocols like ours. 

Another requirement of our system is that of quantum memories for storing and processing entangled states. In principle, we can run our protocol once all required entangled states are shared among users. This may increase the waiting time as well as the required storage/coherence time for memories. For a small-scale demonstration, with a few number of players at short distances from each other, this, can, however, be manageable. Quantum memories such as nitrogen vacancy centers in diamond \cite{Kalb18Reiserer} , or trapped ions \cite{Moehring07Maunz,Schafer07Ballance}, offer long storage times that could be suitable for our protocol. Plus, both these memories offer settings in which high-quality deterministic CNOT gates can be performed. The latter is necessary in order to keep our protocol loss resilient.

In terms of performance, there are two parameters that typically matter: At what rate, we can distribute entangled states among parties, and what would be the quality of the generated entangled state. The rate of entanglement generation is mainly affected by channel loss, but, for moderately short links, this may not be the major obstacle. For instance, if the maximum distance between two players is 50~km, for standard optical fiber channels with 0.2~dB/km loss, we have a channel transmissivity of 0.1. By accounting for a similar efficiency, for other parts of the system, we have a 1\% chance in generating entangled states in every attempt. For a repetition rate of 1~M/s, we can then generate 10,000 entangled links per second, which should be sufficient for a small-scale demonstration. In terms of quality, in our analysis, we have assumed perfect Bell states can be exchanged among users. This is in principle possible if one can use entanglement distillation or error correction techniques. For a simple demonstration, however, it is more likely that we have to accept a bit of error in our system. This error rate would scale with the distance between the shared entangled state versus maximally entangled states, as well as with the number of players. One should also add to that the errors that might arise during the Bell-state measurements. In the end, if the error caused by imperfections in the system is too high, the protocol will abort during its verification stage.

One final note is about the number of Bell states that are needed for attack detection in our protocol. Here, in principle, we are using similar ideas as those used in quantum key distribution (QKD) for detecting eavesdroppers. But, unlike QKD, the ratio L/R, in our case, should be very low to keep the protocol secure. The main reason behind this is that, in any quantum summation protocol, the protocol fails even if only one of the private bits gets revealed. That is, we have no chance to remove the information that has leaked to an eavesdropper once it has happened, whereas, in QKD, one can use privacy amplification to reduced the amount of leaked information about the final key. This seems to be a common issue in all quantum summation protocols and is not specific to our case.

\section{Conclusions} \label{secdiccon}
We proposed a secure multi-party quantum summation protocol based on quantum teleportation, in which a third party (TP), who could be malicious but non-collusive, was involved. The correctness and the security of the protocol were analyzed in detail. 
Our protocol did not require multi-partite entangled states. Only bipartite states (Bell states), Pauli operators and Bell measurement were needed in our protocol. The latter were all required in any teleportation protocol, which would be implicitly used in all other quantum summation protocols as well. By reducing the required resources to those needed for teleportation, we, in effect, proposed the most feasible quantum summation protocol, which could, in principle, be demonstrated, at small scales, using current quantum technologies. A more detailed error analysis is needed to account for the effect of imperfect entanglement distribution and/or operation errors. We will consider these imperfections in our future work.


\vspace{6pt} 



\begin{acknowledgments}
This work is supported by the National Natural Science Foundation of China (Grant Nos.11647140, 61602316, 61872152, 61502179), the Natural Science Foundation of Guangdong Province of China (Grant Nos. 2018A030310147, 2016A030310027, 2014A030310265), Guangdong Program for Special Support of Top-notch Young Professionals (No. 2015TQ01X796), Pearl River Nova Program of Guangzhou (No. 201610010037), the
	Science and Technology Innovation Projects of Shenzhen (No. JCYJ20170818140234295), and the CICAEET fund and the PAPD fund (No. KJR1615).
	Mohsen Razavi acknowledges the support of UK EPSRC Grant EP/M013472/1.
	Cai Zhang is sponsored by the State Scholarship Fund of the China Scholarship Council. All data generated in this paper can be reproduced by the provided methodology.
\end{acknowledgments}


\appendix
\section{Analysis of the Multi-party Quantum Summation} \label{app.analysis}
\unskip
\subsection{Correctness Analysis}
We assume that all participants provide correct private bit strings.
For the convenience of analyzing the correctness of our protocol, we define the relationship between quantum states $ \{|0\rangle,|1\rangle,|+\rangle,|-\rangle\}  $  and classical bits as follows:
\begin{equation}\label{eq.uenc}
E(|\varphi\rangle)=\bigg \lbrace
\begin{tabular}{l}
$0,  \ if \ |\varphi\rangle \in \{|0\rangle,|+\rangle\}$,\\
$1,  \ if \ |\varphi\rangle \in \{|1\rangle,|-\rangle\}$.
\end{tabular}
\end{equation}

Furthermore, if
\begin{equation}\label{eq.uenc2}
|\varphi^{\prime}\rangle= U^{m}|\varphi\rangle,
\end{equation}
where 
$ m \in \{0,1\}$, $ |\varphi\rangle \in \{|0\rangle, |1\rangle,|+\rangle,|-\rangle\} $, $ U=ZX $ and a global phase is ignored, then
\begin{equation}\label{eq.encrec}
E(|\varphi^{\prime}\rangle) = E(|\varphi\rangle) \oplus m .
\end{equation}

In \ref{enu.encoding} of the protocol, $V_{2i}^{i} $ and $ V_{2i+1}^{i}$ ($i=0,1,\ldots,n$) form $L$ ordered Bell states. $V_{0}^{0}$ and $ V_{2n+1}^{n}$ are held by $P_{0}$ and  $V_{2i-1}^{i-1}$ and $V_{2i}^{i}$ ($i=1,2,\ldots,n$) are in $ P_{i} $'s hands. For the $j$-th ($ j=1,2,\ldots,L $) Bell state between 
$V_{2i}^{i} $ and $ V_{2i+1}^{i}$($i=0,1,\ldots,n$), combining with the $j$-th state in quantum memory
$ G^{0}_{T}$, the initial state is 
\begin{equation}\label{eq.corinimul} 
|\zeta_{j}^{0}\rangle=|\varphi_{j}\rangle_{T}\otimes |{\psi^{\prime}}^{0}_{j}\rangle_{01}\otimes  |{\psi^{\prime}}^{1}_{j}\rangle_{23}\otimes \ldots \otimes |{\psi^{\prime}}^{n}_{j}\rangle_{(2n)(2n+1)}.
\end{equation}
Suppose that
\begin{align}
|{\psi^{\prime}}^{0}_{j}\rangle_{01}&=|B_{a_{0}b_{0}}\rangle_{01}^{j},\\  |{\psi^{\prime}}^{1}_{j}\rangle_{23}&=|B_{a_{1}b_{1}}\rangle_{23}^{j},\\
&\ldots, \\
|{\psi^{\prime}}^{n}_{j}\rangle_{(2n)(2n+1)}&=|B_{a_{n}b_{n}}\rangle_{(2n)(2n+1)}^{j},
\end{align}
and
$ P_{i} $  ($i = 1,2,\ldots,n$) performs 
$ U_{i}^{m_{ij}} $ 
($ U_{i}=U=ZX $) on the $j$-th state of $ V_{2i-1}^{i-1}$, the state becomes
\begin{eqnarray}\label{eq.corbri}
\nonumber
|\zeta_{j}^{1}\rangle= &\frac{1}{2^{n+1}}\sum_{x_{0j}\in \{0,1\}}& \sum_{y_{0j}\in \{0,1\}} \sum_{x_{1j}\in \{0,1\}} \sum_{y_{1j}\in \{0,1\}}\ldots \sum_{x_{nj}\in \{0,1\}} \sum_{y_{nj}\in \{0,1\}} \\ \nonumber
&(-1)^{\Sigma_{i=0}^{n} x_{ij}\cdot b_{i}}&
|B_{x_{0j}y_{0j}}\rangle_{T0}^{j}\otimes|B_{x_{1j}y_{1j}}\rangle_{12}^{j}\otimes\ldots\otimes|B_{x_{nj}y_{nj}}\rangle_{(2n-1)(2n)}^{j}\\
&\otimes Z^{\oplus\Sigma_{i=0}^{n}b_{i}\oplus y_{ij}}&X^{\oplus\Sigma_{i=0}^{n}a_{i}\oplus x_{ij}}U^{\oplus\Sigma_{i=1}^{n}m_{ij}}|\varphi_{j}\rangle_{2n+1},
\end{eqnarray}
according to Eqs. (\ref{eq.tel}-\ref{eq.four.last}), and a global phase of the state of quantum memory ($2n+1$) is ignored. 

After $ P_{i}$ ($ i=0,1,\ldots,n$) measures the corresponding states in the Bell basis, obtaining the measurement outcome $ x_{ij}y_{ij}$ ($j=1, 2, \ldots, L$), the state of quantum memory ($2n+1$) collapses to
\begin{equation}\label{eq.fin2}
Z^{\oplus\Sigma_{i=0}^{n}b_{i}\oplus y_{ij}}X^{\oplus\Sigma_{i=0}^{n}a_{i}\oplus x_{ij}}U^{\oplus\Sigma_{i=1}^{n}m_{ij}}|\varphi_{j}\rangle_{2n+1}.
\end{equation}
With the announcement of  $ x_{ij}y_{ij}$  ($ i=1,2,\ldots,n$) provided by $ P_{i}$, $P_{0} $ knowing the initial Bell states can calculate
\begin{eqnarray}\label{eq.sec1}
&\oplus\sum_{i=0}^{n}&a_{i}\oplus x_{ij},\\
\label{eq.sec2}
&\oplus\sum_{i=0}^{n}&b_{i}\oplus y_{ij}.
\end{eqnarray}
Later, $X^{\oplus\Sigma_{i=0}^{n}a_{i}\oplus x_{ij}}Z^{\oplus\Sigma_{i=0}^{n}b_{i}\oplus y_{ij}}$ is performed on quantum memory ($2n+1$).
Consequently, the state of quantum memory ($ 2n+1 $) turns into
\begin{equation}\label{key22}
|\varphi^{\prime}_{j}\rangle_{2n+1}=U^{\oplus\Sigma_{i=1}^{n}m_{ij}}|\varphi_{j}\rangle_{2n+1}.
\end{equation}
After the measurement of quantum memory ($2n+1$) in the same basis as that of quantum memory $T$, $ P_{0} $ gains
\begin{equation}
E(|{\varphi}_{j}\rangle_{T})\oplus (\oplus\Sigma_{i=1}^{n}m_{ij})=E(|{\varphi^{\prime}}_{j}\rangle_{2n+1}),
\end{equation}
and therefore obtains the result
\begin{equation}
\oplus\Sigma_{i=1}^{n}m_{ij}= E(|\varphi_{j}\rangle_{T})\oplus E(|\varphi^{\prime}_{j}\rangle_{2n+1}),
\end{equation}
for the $j$-th bit of the sum modulo $2$ of participants' private bit strings, by using Eqs. (\ref{eq.uenc}-\ref{eq.encrec}). In the end, $P_{0}$ is able to learn about the sum modulo $2$ of participants' private bit strings.

\subsection{Security Analysis}
There exist two types of participant attacks, one from TP($ P_{0} $) and the other from some dishonest participants. We showed earlier how our protocol is secure against attacks by TP. Here we demonstrate how our protocol can be kept secure in the presence of malicious participants. Note that $ n-1 $ dishonest participants can easily steal the honest participant's private bit string if the summation is revealed in public. But if the summation is kept secret in TP's hands, $ n-1 $ dishonest participant cannot obtain anything about the honest participant's private input. Here, we show that our protocol is secure against the collusive attack of $ n-2 $ dishonest participants, which is the maximum possible in this case. 

\textit{Attacks from ($n-2$) dishonest participants (not including $P_{0}$).} If ($ n-2 $) dishonest participants wish to steal the other two honest participants' private bit strings $ M_{p} $ and $ M_{q} $ ($ p < q $), they may employ the states in their hands to get useful information. We consider the $j$-th bit ($ j=1,2,\dots,L $) in $ M_{p} $ and $ M_{q} $ and the corresponding states.

\begin{figure}
	\centering
	\includegraphics[width=0.6\textwidth]{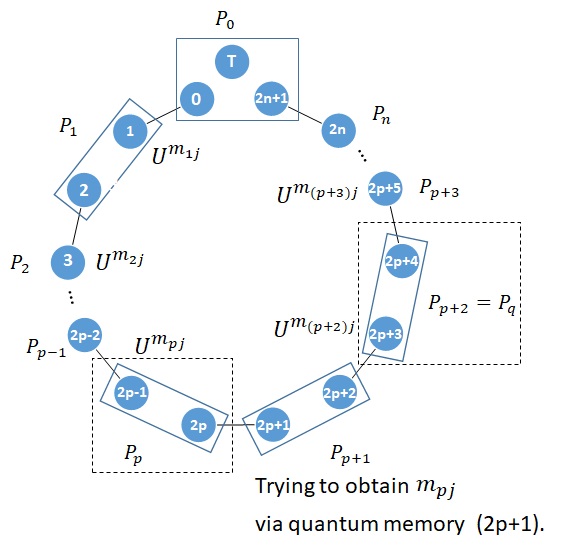}
	\centering
	\caption{Attack by ($n-2$) participants, where $P_{p}$ and $P_{q}$ are honest participants.}\label{fig.dishonest00}
\end{figure}

For $q\neq p+1 $,  we first show how dishonest participants try to learn about $m_{pj}$, as shown in Figure \ref{fig.dishonest00}. In this case, $ P_{p+1} $ does not apply unitary operation on quantum memory ($2p+1$) and Bell-state measurement on quantum memories ($2p+1$) and ($2p+2$). After the private input encoding stage \ref{enu.encoding}, the state of quantum memory $ T $ and quantum memories $ 0\sim(2p+1) $ will be
\begin{eqnarray}\label{eq.corbri22}
\nonumber
|\zeta_{j}^{1}\rangle= &\frac{1}{2^{p}}\sum_{x_{0}\in \{0,1\}}& \sum_{y_{0}\in \{0,1\}} \sum_{x_{1}\in \{0,1\}} \sum_{y_{1}\in \{0,1\}}\ldots \sum_{x_{p}\in \{0,1\}} \sum_{y_{p}\in \{0,1\}} \\ \nonumber
&(-1)^{\Sigma_{k=0}^{p} x_{k}\cdot b_{k}}&
|B_{x_{0}y_{0}}\rangle_{T0}^{j}\otimes|B_{x_{1}y_{1}}\rangle_{12}^{j}\otimes\ldots\otimes|B_{x_{p}y_{p}}\rangle_{(2p-1)(2p)}^{j}\\
&\otimes Z^{\oplus\Sigma_{k=0}^{p}b_{k}\oplus y_{k}}&X^{\oplus\Sigma_{k=0}^{p}a_{k}\oplus x_{k}}U^{\oplus\Sigma_{k=1}^{p}m_{kj}}|\varphi_{j}\rangle_{2p+1},
\end{eqnarray}
where the $ j $-th state in quantum memory $T$ is $|\varphi_{j}\rangle_{T}$ and the $ j $-th Bell state shared between $ P_{s} $ and $ P_{s+1} $ ($s = 0,1,\ldots,p$) is $ |B_{a_{s}b_{s}}\rangle_{(2s)(2s+1)}^{j}$. The dishonest participants try to get $m_{pj}$ from quantum memory ($2p+1$). However, they will fail.

From Eq. (\ref{eq.corbri22}), we can see that if $ P_{p+1} $ knows $m_{sj}$ ($ s=1, 2, \ldots,p-1 $)
, the basis of $ |\varphi_{j}\rangle_{T} $ and ($a_{r},b_{r}$) ($ r=0,1,\ldots,p $) (the information about the initial Bell states), she can first apply the right correction on quantum memory ($2p+1$) and measure it in the right basis. According to $m_{sj}$ ($ s=1, 2, \ldots,p-1$), she can then obtain $m_{pj}$. But she cannot do that. Even though $ P_{p+1} $ knows $m_{sj}$ ($ s=0,1,\ldots,p-1$)  with the assistance of $ P_{s} $ and the measurement results ($x_{0}y_{0}, x_{1}y_{1}, \ldots , x_{p}y_{p}$), she knows nothing about the basis of $ |\varphi_{j}\rangle_{T} $ and ($a_{r},b_{r}$) that are kept secret by $P_{0}$. Thus, she cannot perform the right correction on quantum memory ($2p+1$) and measure it in the right basis. Finally, she fails to obtain $m_{pj}$, let alone $ M_{p} $. Similarly, they cannot learn about $ M_{q} $.

For $q=p+1$, they may use a similar method as in the above case to take $M_{p}$ and $M_{q}$. Namely, $P_{p+2}$ does nothing on quantum memory ($2p+3$) and skips Bell-state measurement on the corresponding state.  In this case. the dishonest participants cannot even get the $m_{pj}\oplus m_{(p+1)j}$. Therefore, the privacy of $ M_{p} $ and $ M_{q} $ is preserved.

For any two Bell states $|B_{xy}\rangle_{12}$ and $|B_{ab}\rangle_{34}$, if quantum memories $ 2 $ and $ 3 $ are measured in the Bell basis and the measurement outcome $ |B_{km}\rangle_{23}$ is obtained, the state of quantum memories $ 1 $
and $ 4 $ then collapses to $ |B_{xy\oplus ab \oplus km} \rangle_{14}$ due to the Bell entanglement swapping property.

\begin{figure}
	\centering
	\includegraphics[width=0.7\textwidth]{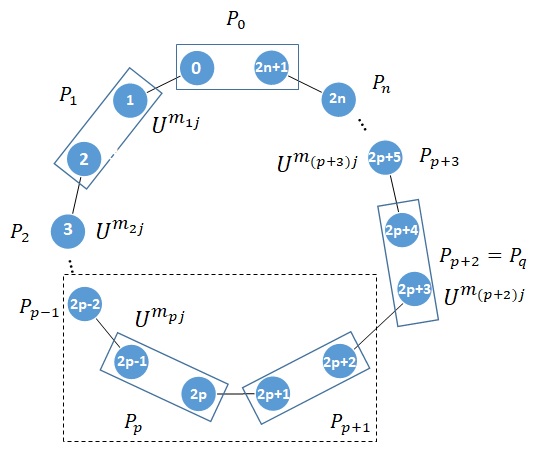}
	\centering
	\caption{Entanglement swapping attack by ($n-2$) participants, where $P_{p}$ and $P_{q}$ are honest participants.}\label{fig.dishonest_engtangled_00}
\end{figure}

The dishonest participants may also start an attack based on the entanglement swapping property. For the case of $q\neq p+1$, as shown in the dash box in Figure \ref{fig.dishonest_engtangled_00}, the $ j $-th Bell state shared between $P_{p-1} $ and $ P_{p} $ and  that shared between $ P_{p} $ and $ P_{p+1} $ are 
$ |B_{a_{p-1}b_{p-1}}\rangle^{j}_{(2p-2)(2p-1)}$ and $ |B_{a_{p}b_{p}}\rangle^{j}_{(2p)(2p+1)}$, respectively. After $ P_{p} $ performs 
$ U_{p}^{m_{pj}} (U_{p}=ZX)$ on quantum memory ($ 2p-1 $) and then measures quantum memories ($2p-1$) and ($ 2p $) in the Bell basis, obtaining the measurement outcome $ |B_{x_{p}y_{p}}\rangle_{(2p-1)(2p)}^{j} $, the state of quantum memories ($ 2p-2 $) and ($ 2p+1 $) becomes
\begin{equation}\label{eq.final}
(I\otimes U_{p}^{m_{pj}})|B_{(a_{p-1}b_{p-1})\oplus (a_{p}b_{p})\oplus (x_{p}y_{p})}\rangle_{(2p-2)(2p+1)}^{j},
\end{equation} 
due to the property of entanglement swapping. $P_{p+1}$ skips the private input encoding stage, instead she can collaborate with $P_{p-1}$ to measure quantum memories ($2p-2$) and ($2p+1$) in the Bell basis. Can the dishonest participants find out $ U_{p}^{m_{pj}} $ performed by $ P_{p} $ to steal $ m_{pj} $?
The answer is no. Although $ P_{p-1} $ and $ P_{p+1} $ can measure quantum memories ($ 2p-2 $) and  ($ 2p+1 $) in the Bell basis and get $ x_{p}y_{p} $ after $ P_{p} $'s announcement, they have to know $ a_{p-1}b_{p-1} $ and $ a_{p}b_{p} $ to derive $ U_{p}^{m_{pj}} $, but this information is unknown to them. For the case of $ q=p+1 $, the analysis is similar. Therefore, this attack is also invalid to our protocol.


\end{document}